\documentclass[10pt,a4paper]{article}
\usepackage{hyperref}
\usepackage[numbers]{natbib}
\usepackage{rotating}
\usepackage{graphicx}

\def\e{\rm {e}}
\def\G{\mathcal{G}}
\def\H{\mathcal{H}}
 
\title{Fractional Cable Model for Signal Conduction in Spiny Neuronal Dendrites}

\author{Silvia  Vitali \\ email \href{mailto:silvia.vitali4@unibo.it}{silvia.vitali4@unibo.it} 
   \and Francesco Mainardi \\ email \href{mailto:francesco.mainardi@bo.infn.it}{francesco.mainardi@bo.infn.it} }
\begin{document}

\maketitle

\begin{abstract}
The cable model is widely used in several fields of science to describe the propagation of signals. 
 A relevant  medical and biological example is 
 the anomalous subdiffusion in spiny neuronal dendrites observed in several studies of the last decade.  
 Anomalous subdiffusion can be modelled in several ways introducing  
 some fractional component into the classical cable model. The Chauchy problem associated to these kind of models 
 has been investigated by many authors, but up to our knowledge an explicit solution for the signalling problem 
 has not yet been published.
Here we propose how this solution can be derived applying the generalized convolution theorem (known as  Efros theorem)  for Laplace transforms.

The fractional cable model considered in this paper is defined by replacing the first order time derivative
 with a  fractional derivative of order $\alpha\in(0,1)$ of Caputo type.
 The signalling problem is solved for any input function applied to the accessible end  of a  semi-infinite cable, which satisfies the requirements of the Efros theorem. The solutions corresponding to the  simple cases
  of impulsive and step inputs are explicitly calculated in integral form containing  Wright functions.  Thanks to 
 the variability of the parameter $\alpha$, the corresponding  solutions    are expected  to  adapt to the qualitative  behaviour of the membrane potential observed in experiments better than in the standard case $\alpha=1$.
 
\end{abstract}


    \maketitle

\section*{Introduction}
	
The one dimensional cable equation has been used from a long time to describe the spatial 
and the temporal dependence of trans-membrane potential $V_m(x,t)$ along the
axial $x$ direction of a cylindrical nerve cell segment. 
It can be derived directly from the Nernst-Planck equation for 
electro-diffusive motion of ions.
The resulting differential equation for the trans-membrane potential 
takes the form of a standard diffusion equation with a shift:
\begin{equation}\label{linear}
 \lambda^2\frac{\partial^2V_m(x,t)}{\partial x^2}-\gamma \frac{\partial V_m(x,t)}{\partial t}-V_m(x,t)=0\,,
\end{equation}
where $\lambda$ and $\gamma$ are space and time constants related  to the membrane resistance and capacitance per unit length, see e.g.  \cite{Magin}.
For simplicity in the rest of this work, following \cite{Magin},
we will use the dimensionless scaled variables $X=x/\lambda$ and $T=t/\gamma$,  
so that we consider the equation 
\begin{equation}\label{linear-scaled}
 \frac{\partial^2V_m(X,T)}{\partial X^2}-
  \frac{\partial V_m(X,T)}{\partial T}-V_m(X,T)=0\,.
\end{equation}
In signalling problems the cable  is considered of semi-infinite length
($0\le X<\infty$, initially quiescent for $T<0$ and excited for $T\ge 0$ at the accessible end ($X=0$) with a given input in  membrane potential $V_m(0,T) = g(t)$.      
Fundamental problems are the cases of an impulsive input $g(t)= \delta(t)$
and of a unit step input  $g(t) = \theta(t)$ where $\delta(t)$  and $\theta(t)$ denote the Dirac  and the Heaviside functions, respectively. The  solutions
corresponding to these inputs read in our notation
\begin{equation}\label{linsol}
 \G_{s}(X,T)=\frac{X}{\sqrt{4\pi T^3}}\e^{-(\frac{X^2}{4T}+T)}\,,
\end{equation}
 and 
\begin{equation}\label{linsol-step}
 \H_{s}(X,T)=\int_0^T \G_{s}(X,T')\, dT'\,.
\end{equation}
We  refer to $\G_s$ to as the fundamental solution or the Green function
for the signalling problem  of the  (standard)  cable equation   (\ref{linear-scaled}),  
whereas to $\H_s$ to as the step response.
As known, the Green function is used in the time convolution integral to represent the solution corresponding to any given input $g(T)$ as follows
  \begin{equation}\label{linsol-general}
 V_m(X,T) =\int_0^T g(T-T')\, \G_s(X,T')\, dT'\,.
\end{equation}
The spatial variance associated to this model is known to evolve linearly in time,
while it has been observed that the spatial variance of an inert tracer in spiny Purkenje cell dendrites
evolves as a sub-linear power law of time,
as spines may trap and release diffusing molecules, and the diffusion with smaller values of the power exponent is associated
to higher spine density \cite{Fide-2006}.

To model anomalous sub-diffusion we substitute the first-order time derivative 
in Eq.(\ref{linear-scaled}) with a  fractional time derivative of Caputo type 
 \cite{GorMai-CISM97}, \cite{Podlubny-BOOK99} of order $\alpha \in (0,1)$:
\begin{equation}\label{frac}
 \frac{\partial^2 V_m(X,T)}{\partial X^2}-\frac{\partial^\alpha V_m(X,T)}{\partial T^\alpha}- V_m(X,T)=0\,.
\end{equation}
This kind of model is a simple extension to fractional behaviour of the Neuronal Cable Model
from a mathematical point of view
and it turns to be in some way equivalent to the equation developed in a relevant study \cite{Henry2008}, which has been derived
from a modified Nernst-Planck equation, with diffusion constant replaced by fractional derivatives of  Riemann-Liouville type.
Other studies consider similar approaches \cite{Henry2008},\cite{T-2009},\cite{Langlands-2011},\cite{Liu-2011},\cite{Moaddy}. 
We will see that beside the apparent simplicity 
our approach allows to reproduce at least qualitatively the main characteristics observed in experiments \cite{Nimchinsky-2002},
\cite{Jacobs-1997},\cite{Duan-2003},\cite{Motohir-1980}. 
Further generalizations
of this model introducing a second fractional time derivative to the shift term could be analysed in future, to refine
the biological relevance of the model.

	
\section*{Solution of the signalling problem}
Applying the Laplace transform to Eq. (\ref{frac})  with the  boundary conditions
required by the signalling problem, that is    $V_m(X,0^+)=0$, $V_m(0,T)=g(T)$,  we have:
\begin{equation}
 (s^\alpha+1)\widetilde{V_m}(X,s)- \frac{\partial^2 \widetilde{V_m}(X,s)}{\partial X^2}=0,
\end{equation}
which is a second order equation in the variable $X$ with solution:
\begin{equation}\label{LT-cable}
 \widetilde{V_m}(X,s)= \widetilde{g}(s) e^{ - \sqrt{(s^\alpha + 1)} \cdot X} .
\end{equation}
The inversion of the Laplace transform now requires special effort with respect to the case where the term $V_m(X, T)$ is not present  in  Eq. (\ref{frac}), that is for 
 \begin{equation}\label{frac-standard}
 \frac{\partial^2 V^*_m(X,T)}{\partial X^2}-\frac{\partial^\alpha V^*_m(X,T)}{\partial T^\alpha}=0\,.
\end{equation}
Indeed for Eq. (\ref{frac-standard}), known as the time-fractional diffusion equation, the solutions of the corresponding Cauchy and signalling problems have been found in the 1990's by Mainardi in terms of 2 auxiliary Wright functions (of the second type)
\cite{Mainardi-CSF96, Mainardi-CISM97}. Specifically for the signalling problem
the general solution provided by Mainardi in integral convolution form reads
\begin{equation}\label{sol-frac-standard-W}
 V^*_m(X,T)
 =\int_0^T g(T-T')\, \G^*_{\alpha, s}(X, T')\, dT'\,,\quad
\G^*_{\alpha, s}(X,T) = 
\frac{1}{T} W_{-{\alpha/2},0}\left(-X/T^{\alpha/2}\right) \,,
\end{equation}
where 
 $\G^*_{\alpha, s}(X,T) $ denotes  the Green function of the signalling problem of the fractional time diffusion equation (\ref{frac-standard}) and  
$W_{-{\alpha/2},0}(\cdot)$ is a particular case of the transcendental function   known as Wright function
\begin{equation} \label{Wright-function}
 W_{\lambda, \mu }(z) := \sum_{n=0}^{\infty}
 \frac{ z^n}{  n!\, \Gamma[\lambda  n + \mu ]}\,,\quad
 \lambda>-1, \; \mu\ge 0\,.
 \end{equation}
 This  function, entire in the complex plane, 
 is extensively discussed in the Appendix F of 
 Mainardi's book \cite{Mainardi-BOOK10} where the interested reader can find the 
 following relevant Laplace transform pairs, rigorously  derived by 
 Stankovi{\'c}\cite{Stankovic-WRIGHT70}:
 \begin{equation}\label{Stankovic}
  t^{\mu-1} \, W_{-\nu, \mu} \left(x/t^\nu\right)
\, \div \, s^{-\mu} \,\exp{\left(-x s^\nu\right)} \ 
  \,,
  \quad 0\le \nu<1\,, \; \mu>0\,.
  \end{equation}
  Here we have   adopted an obvious notation to denote the juxtaposition of a locally integrable  function of time $t$ with its Laplace transform in $s$ with $x$ a positive parameter.
 It is worth to  recall  the distinction of the Wright functions in first type 
 ($\lambda \ge 0$)
  and second type ($-1<\lambda\le 0$)  and, among the latter ones, the relevance of the two auxiliary functions introduced in \cite{Mainardi-CSF96}:
 \begin{equation} \label{F-M}
 F_\nu(z) =  W_{-\nu, 0}(-z)\,, 
  \quad M_\nu(z) =    W_{-\nu, 1-\nu}(-z)\,, \quad 0<\nu<1\,,
  \end{equation}
  inter-related  as $F_\nu(z) = \nu z M_\nu(z) $.
 Indeed the relevance of both the Wright functions
has been outlined by several authors  in diffusion and stochastic processes. 
Particular attention is due  to the   $M$-Wright function (also referred to as   the Mainardi function in  \cite{Podlubny-BOOK99}) 
  that, since  for $\nu=1/2$ reduces to $\exp{(-z^2/4)}/ \sqrt{\pi}$,
is considered  a suitable generalization of the Gaussian 
density, see \cite{Pagnini-FCAA13} and references therein.

Then the Green function for the signalling problem of the time fractional diffusion equation (\ref{frac-standard}) can be written in the original form provided in 
\cite{Mainardi-CSF96} 
 as
 \begin{equation}\label{sol-frac-standard-F-M}
 \G^*_{\alpha, s}(X,T) = 
\frac{1}{T} F_{\alpha/2}\left(X/T^{\alpha/2}\right) =
\frac{\alpha}{2}\, \frac{X}{ T^{\alpha/2+1}}\,
M_{\alpha/2}\left(X/T^{\alpha/2}\right)\,,
\end{equation}
where the superscript $*$  is added  to distinguish the time fractional diffusion equation from our  fractional cable equation, both depending on the order 
$\alpha\in (0,1)$.
   
Because of the shift constant in the square root of the Laplace transform 
in Eq.(\ref{LT-cable}),  the inversion is no longer  straightforward with the Wright functions as it is in the time fractional diffusion equation (\ref{frac-standard}). 
Consequently,  we 
have overcome this difficulty recurring to the application of the Efros  theorem that generalizes the well known convolution theorem for Laplace transforms. For sake of  convenience let us  hereafter recall this theorem, usually not so well-known  in the literature. 

 The Efros theorem \cite{Graf} states  that if we can write a Laplace transform 
 $\widetilde{f}(s)$ as: 
\begin{equation}
\widetilde{f}(s) = \phi(s) \cdot \widetilde{F}(\psi(s)) ,
\end{equation}
where the function $\widetilde{F}(s)$ has a known inverse Laplace transform $F(T)$,
the inverse Laplace transform is: 
\begin{equation}
 f(T)= \int_0^\infty F(\tau)G(\tau,T)d\tau
\end{equation}
where:
\begin{equation}
G(\tau;T)\div \widetilde{G}(\tau,s)= \phi(s) e^{-\tau \psi(s)}
\end{equation}
For the solution (\ref{LT-cable})
of our signalling problem  
  we thus have:
\begin{equation}
 \phi(s)=\widetilde{g}(s),\quad \psi(s)=s^\alpha\,,
\end{equation}
and
\begin{equation}
\widetilde{F}(s)|_{X}  =\e^{-X\sqrt{s+1}}\,.
\end{equation}
Then, having $\widetilde{G}(\tau,s)=
\widetilde{g}(s)\, \e^{-\tau s^\alpha}$, thanks to the standard convolution theorem of Laplace transforms, we obtain:
\begin{equation}
 G(\tau,T)=\int_0^T \frac{g(T-T')}{T'}W_{-\alpha,0}(-\tau/T'^\alpha)dT'
\end{equation}
where $W_{-\alpha,0}$ is the  F-Wright function, and 
\begin{equation}
F(T)|_{X}= \frac{X}{\sqrt{4\pi T^3}}e^{-(\frac{X^2}{4T}+T)}
\end{equation}
is the solution (\ref{linsol})
 of the standard  cable equation (\ref{linear-scaled}).

Then, the general solution of our signalling problem can be written in terms of known functions:
\begin{eqnarray}\label{general solution}
 V_m(X,T)&&=\int_0^\infty \frac{X}{\sqrt{4\pi \tau^3}}\,\e^{-(\frac{X^2}{4\tau}+\tau)}
 \left[\int_0^T \frac{g(T-{T'})}{{T'}}W_{-\alpha,0}(-\tau/{T'}^\alpha)d{T'}\right]\, d\tau
 \nonumber\\
 &&=\int_0^T g(T-{T'})\left [\int_0^\infty \frac{X}{\sqrt{4\pi \tau^3}}
 {\rm e}^{-(\frac{X^2}{4\tau}+\tau)}
\frac{1}{{T'}} W_{-\alpha,0}(-\tau/{T'}^\alpha) \,d\tau \right]\,d{T'}\,.
\end{eqnarray}

\subsubsection*{Example 1: $g(T)=\delta(T)$}
Substituting $g(T)=\delta(T)$ in the general solution 
(\ref{general solution})
we obtain the Green function for the fractional model (\ref{frac}):
\begin{eqnarray}
V_m(X,T) := 
 \G_{\alpha,s}(X,T)&&=\int_0^\infty \G_s(X,\tau)
\frac{1}{T} W_{-\alpha,0}(-\tau/T^\alpha) d\tau\nonumber\\
&&=\int_0^\infty \G_s(X,\tau)
\frac{1}{T} F_{\alpha}(\frac{\tau}{T^\alpha}) d\tau
\end{eqnarray}
This solution is plotted versus $X$ for $T=0.1$ and $T=100$ and versus $T$ 
 for $X=1$ in Fig.1.

\begin{center}
\includegraphics[width=6cm]{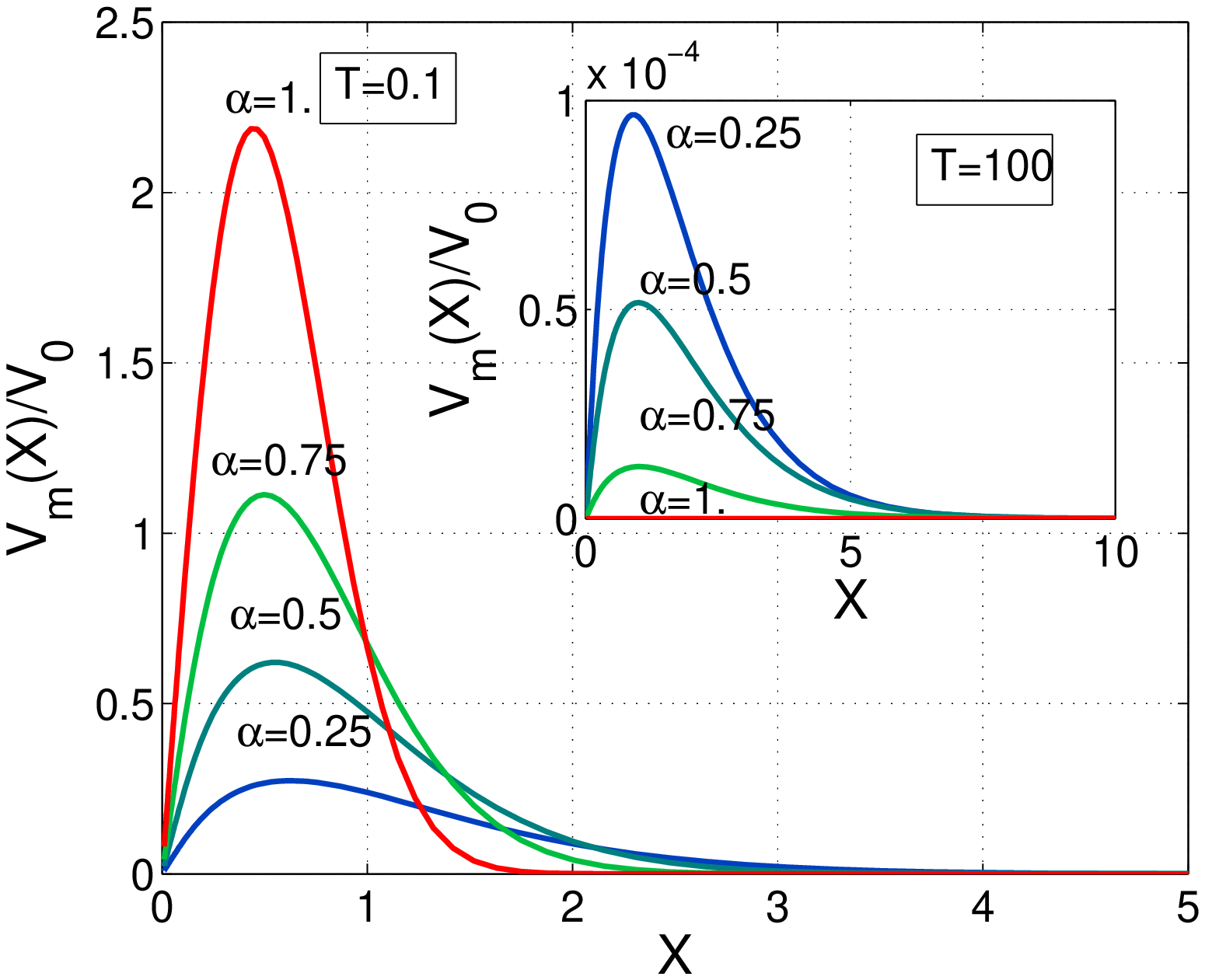}
\includegraphics[width=6cm]{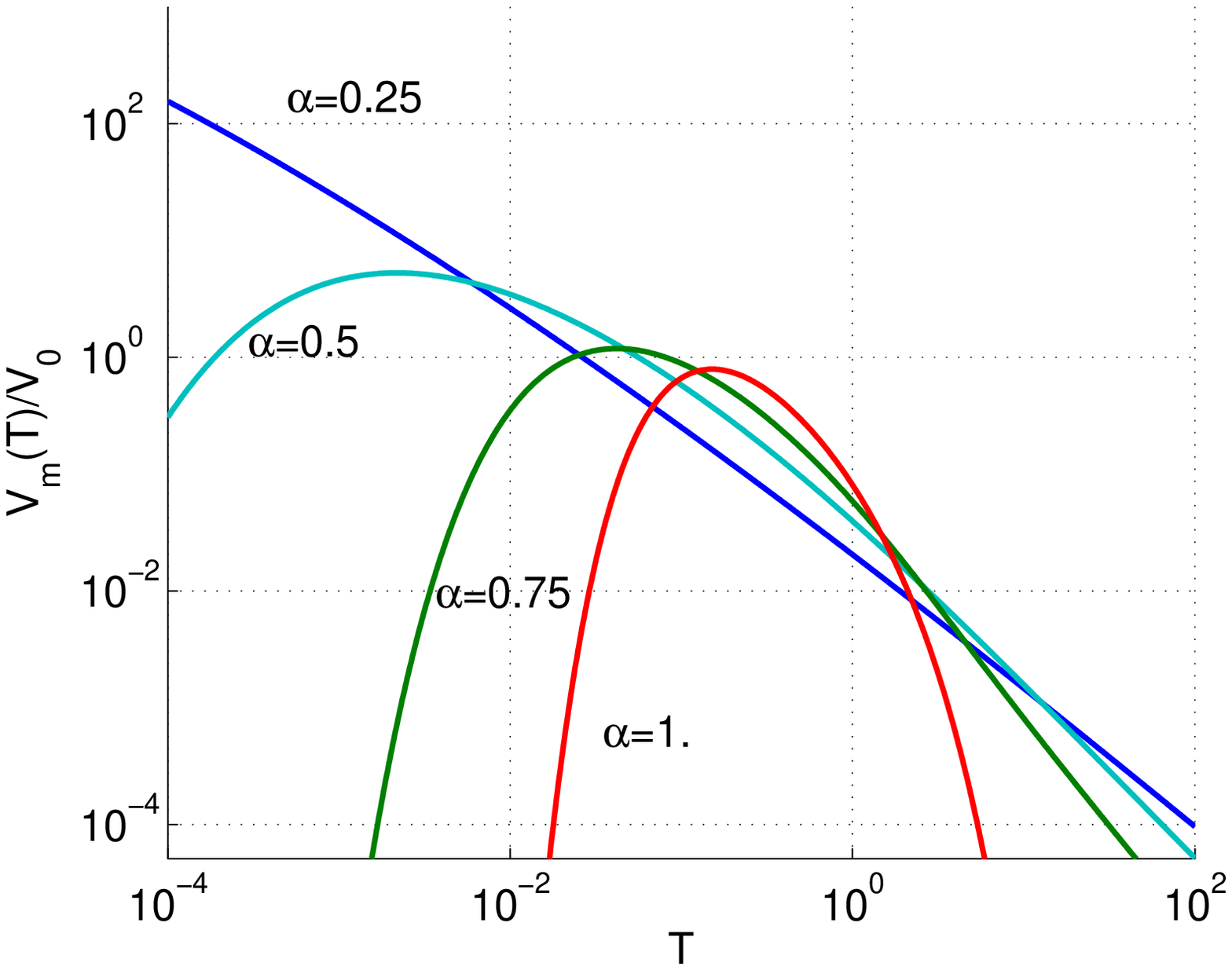}
\end{center}
\vskip 0.1truecm
Fig.1: Plots of $V_m(X,T)$ versus $X$ in $T=0.1$ and $T=100.$ in the inset (left) and plots of $V_m(X,T)$ versus $T$ in $X=1.$ (right), for $g(T)=\delta(T)$.
\vskip 0.5truecm

\subsubsection*{Example 2: $g(T)=\theta(T)$}

When $g(T)=\theta(T)$
 we obtain the step response of our fractional cable equation :
\begin{eqnarray}
 V_m(X,T) := 
 \H_{\alpha,s}(X,T) 
 &&=\int_0^\infty \frac{X}{\sqrt{4\pi \tau^3}}\, \e^{-(\frac{X^2}{4\tau}+\tau)}
 \left[\int_0^T \frac{1}{{T'}}W_{-\alpha,0}(-\tau/{T'}^\alpha)d{T'}\right] \, d\tau\,.
\end{eqnarray}
After some manipulations including  the change of variable $z=\tau/{T'}^\alpha$ and
integrating by parts after using  the recurrence relation
of Wright functions:
\begin{equation}
\frac{dW_{\lambda,\mu}(z)}{dz}=W_{\lambda,\lambda+\mu}(z)
\end{equation}
and the relation between the auxiliary functions:
$ F_\nu(z)=\nu z \,M_\nu(z)$
we may rewrite the step-response solution as:
\begin{equation}\label{teta}
  V_m(X,T):= 
 \H_{\alpha,s}(X,T) 
  = \int_0^\infty \H_s(X,\tau) \cdot\frac{1}{T^\alpha}M_\alpha(\frac{\tau}{T^\alpha})d\tau
\end{equation}
This solution is plotted versus $X$ for $T=1$ and versus $T$ 
 for $X=1$ in Fig.2.
\begin{center}
\includegraphics[width=6cm]{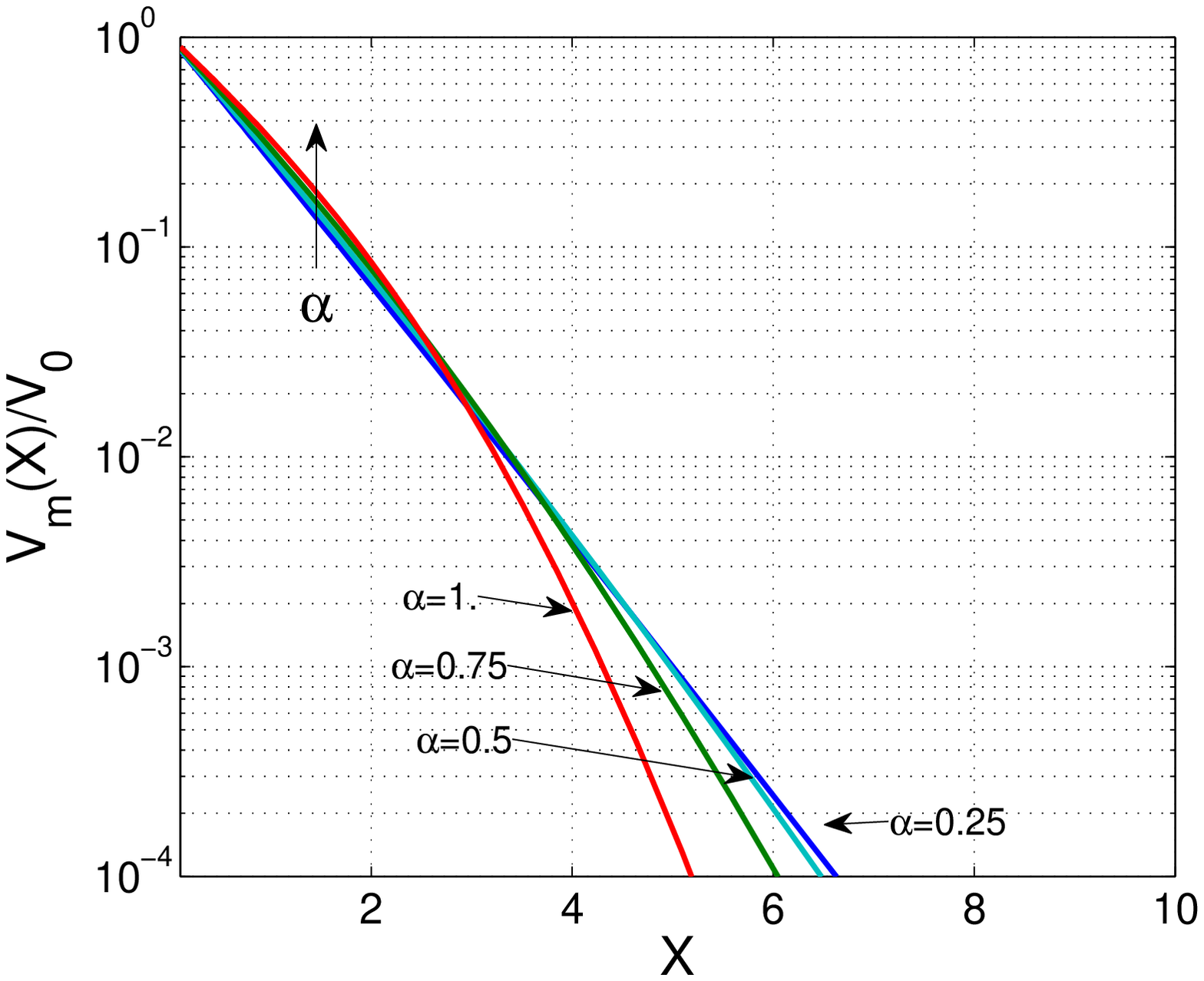}
\includegraphics[width=6cm]{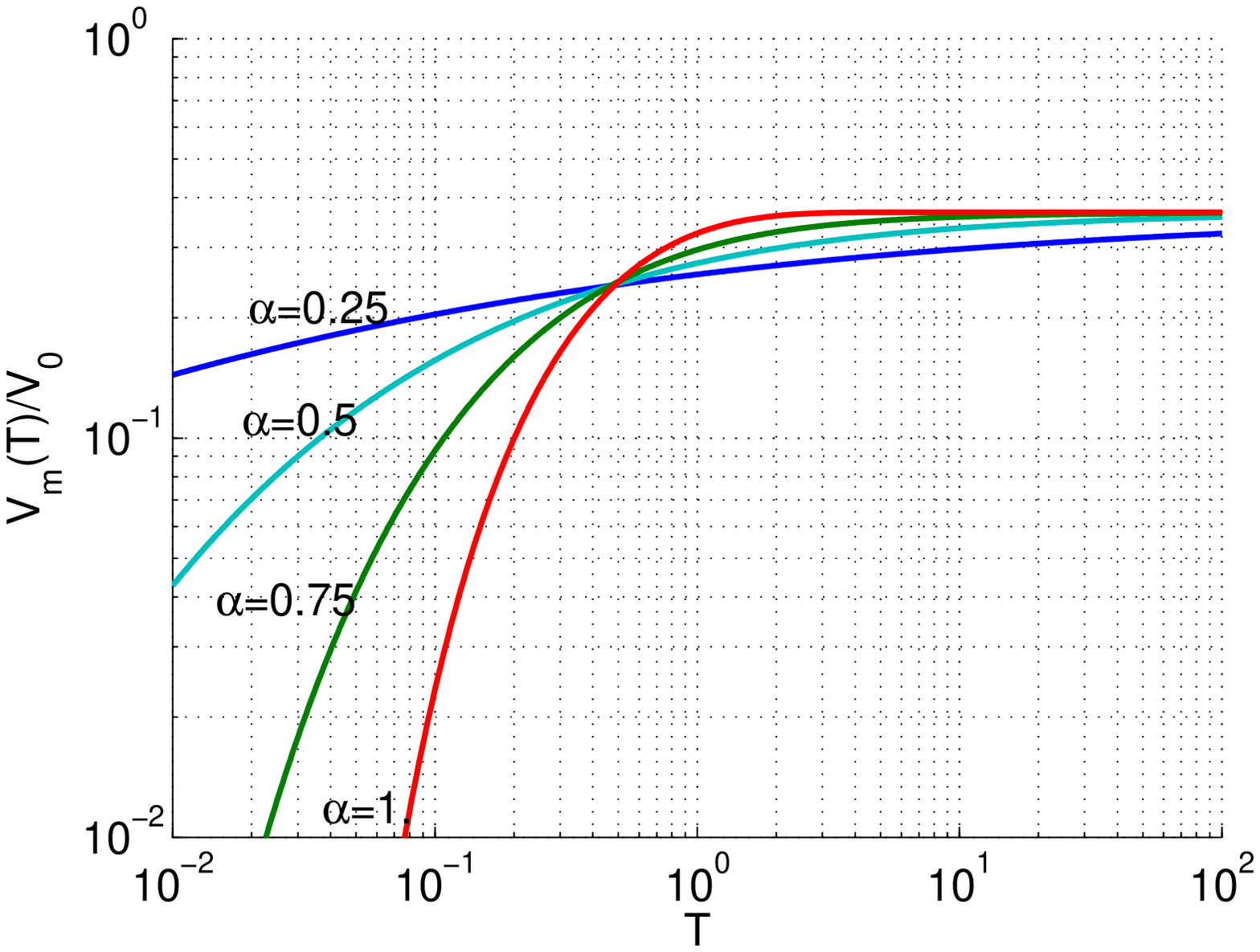}
\end{center}
\vskip 0.1truecm
Fig.2: Plots of $V_m(X,T)$ versus $X$ in $T=1.$ (left) and plots of $V_m(X,T)$ versus $T$ in $X=1.$ (right), for $g(T)=\theta(T)$.
\vskip 0.5truecm

The plotted solutions were  computed without any particular difficulty except the careful  truncation of the series defining the auxiliary $F$ and $M$ functions in our routines.

\section*{Biological interpretation}
Fractional cable models are used to describe subthreshold potentials, or passive potentials,
associated to dendritic processes in neurons. 
The travelling potential is summed up at the soma and the cell produces an action potential when a threshold is exceeded.

It has been observed in \cite{Fide-2006} that diffusion results more anomalous, i.e. the fractional exponent $\alpha$ decreases,
with increasing spine density. Decreasing spine density is characteristic of aging \cite{Jacobs-1997},\cite{Duan-2003}, pathologies
as neurological disorders \cite{Nimchinsky-2002} and Down's sindrome \cite{Motohir-1980}. It has been suggested that increasing spine density should serve 
to compensate time delay of postsynaptic potentials along dendrites and to reduce their long time temporal attenuation \cite{Henry2008}.

Looking at our plotted solutions for the fractional cable equation  short and long space and time behaviour can be distinguished about the evolution of the sub-diffusion process. 

When an impulsive potential  is applied  at the accessible end  it can be noted from Fig.1  that peak high decreases more rapidly with decreasing 
$\alpha$ at early times, viceversa is less suppressed at longer times, and the cross over time increases with decreasing $\alpha$.
Looking at the potential versus time it can also be noted that potential functions
associated to lower $\alpha$ last for longer time at 
appreciable intensity and arrive faster at early times with respect to the normal diffusion case.

By the way, when a constant potential is applied  at the accessible end  we note from Fig.2 that  the exponential suppression of the potential along the dendrite is reduced for high $X$ values with respect to 
normal diffusion. Instead for small $X$ the potential results just slightly more suppressed in the  sub-diffusion process.

\section*{Conclusions}
The presented fractional cable model satisfies the main biological features of the dendritic cell signalling problem.
With respect to models solved as Cauchy problem, our approach could include specific
time dependent boundary conditions, which will allow to reconstruct with accuracy 
the expected signal at the soma if the model will result capable to predict real data behaviour. 
Furthermore the 
solutions can be computed directly, i.e.
calculating the integral associated, as well as by Laplace transform inversion 
\cite{Abate} without any remarkable issue.

\section*{Aknowledgements}
The work of 
F. M.  has been carried out in the framework of the activities of the
 National Group of Mathematical Physics (INdAM-GNFM).
The  authors are indebted to the 
Interdepartmental Center  "Luigi Galvani" for integrated studies of Bioinformatics,
Biophysics and Biocomplexity of the University of Bologna for partial support.


\providecommand{\noopsort}[1]{}\providecommand{\singleletter}[1]{#1}%


\begin{thebibliography}{10}

\bibitem{Abate}
J.~Abate and W.~Ward.
\newblock A unified framework for numerically inverting laplace transforms.
\newblock {\em INFORMS J. on Computing}, 18(4):408--421, January 2006.

\bibitem{Duan-2003}
H.~Duan.
\newblock Age-related dendritic and spine changes in corticocortically
  projecting neurons in {M}acaque monkeys.
\newblock {\em Cerebral Cortex}, 13, 09 2003.

\bibitem{GorMai-CISM97}
R.~Gorenflo and F.~Mainardi.
\newblock {\em Fractional Calculus: Integral and Differential Equations of
  Fractional Order}, pages 223--276.
\newblock CISM Courses and Lecture Notes, Vol. 378. Springer-Verlag, Wien,
  1997.

\bibitem{Graf}
U.~Graf.
\newblock {\em Applied Laplace Transforms and z-Transforms for Scientists and
  Engineers}.
\newblock Springer, 2004.

\bibitem{Henry2008}
B.I. Henry, T.A.M. Langlands, and S.L. Wearne.
\newblock Fractional cable models for spiny neuronal dendrities.
\newblock {\em Phys. Rev. Lett}, 100:128103/1--3, 2008.

\bibitem{Jacobs-1997}
Bob Jacobs, Lori Driscoll, and Matthew Schall.
\newblock Life-span dendritic and spine changes in areas 10 and 18 of human
  cortex: A quantitative golgi study.
\newblock {\em The Journal of Comparative Neurology}, 386, 10 1997.

\bibitem{T-2009}
T.A.M. Langlands, B~I. Henry, and S.L. Wearne.
\newblock Fractional cable equation models for anomalous electrodiffusion in
  nerve cells: infinite domain solutions.
\newblock {\em Journal of Mathematical Biology}, 59, 2009.

\bibitem{Langlands-2011}
T.A.M. Langlands, B.I. Henry, and S.L. Wearne.
\newblock Fractional cable equation models for anomalous electrodiffusion in
  nerve cells: Finite domain solutions.
\newblock {\em SIAM Journal on Applied Mathematics}, 71, 2011.

\bibitem{Liu-2011}
Fawang Liu, Qianqian Yang, and Ian Turner.
\newblock Two new implicit numerical methods for the fractional cable equation.
\newblock {\em Journal of Computational and Nonlinear Dynamics}, 6, 2011.

\bibitem{Magin}
R.L Magin.
\newblock {\em Fractional Calculus in Bioengineering}.
\newblock Begell House Publishers, 2006.

\bibitem{Mainardi-CSF96}
F.~Mainardi.
\newblock Fractional relaxation-oscillation and fractional diffusion-wave
  phenomena.
\newblock {\em Chaos, Solitons and Fractals}, 7:1461--1477, 1996.

\bibitem{Mainardi-CISM97}
F.~Mainardi.
\newblock {\em Fractional Calculus: Some Basic problems in Continuum and
  Statistical Mechanics}, pages 291--348.
\newblock CISM Courses and Lecture Notes, Vol. 378. Springer-Verlag, Wien,
  1997.

\bibitem{Mainardi-BOOK10}
F.~Mainardi.
\newblock {\em Fractional Calculus and Waves in Linear Viscoelasticity}.
\newblock Imperial College Press, London, 1st edition, 2010.

\bibitem{Moaddy}
K.~Moaddy, A.~G. Radwan, K.~N. Salama, S.~Momani, and I.~Hashim.
\newblock The fractional-order modeling and synchronization of electrically
  coupled neuron systems.
\newblock {\em Comput. Math. Appl.}, 64(10):3329--3339, November 2012.

\bibitem{Nimchinsky-2002}
Esther~A. Nimchinsky, Bernardo~L. Sabatini, and Karel Svoboda.
\newblock Sructure and function of dendritic spines.
\newblock {\em Annual Review of Physiology}, 64, 03 2002.

\bibitem{Pagnini-FCAA13}
G.~Pagnini.
\newblock The {M}-{W}right function as a generalization of the {G}aussian
  density for fractional diffusion processes.
\newblock {\em Fract. Calc. Appl. Anal}, 16, 2013.

\bibitem{Podlubny-BOOK99}
I.~Podlubny.
\newblock {\em Fractional Differential Equations}.
\newblock Mathematics in Science and Engineering 198. Academic Press, San
  Diego, 1st edition, 1999.

\bibitem{Motohir-1980}
M. Suetsugu and P. Mehraein.
\newblock Spine distribution along the apical dendrites of the pyramidal
  neurons in {D}own's syndrome.
\newblock {\em Acta Neuropathologica}, 50, 1980.

\bibitem{Stankovic-WRIGHT70}
B.~Stankovi\`c.
\newblock On the function of {E}.{M}. {W}right.
\newblock {\em Publ. de l'InstitutMath\`ematique, Beograd, Nouvelle S\`er.},
  10:113--124.

\bibitem{Fide-2006}
F. Santamaria, S. Wils, E.~De Schutter, and George~J. Augustine.
\newblock Anomalous diffusion in purkinje cell dendrites caused by spines.
\newblock {\em Neuron}, 52, 2006.

\end{thebibliography}
\end{document}